\documentclass[%
superscriptaddress,
showpacs,preprintnumbers,
 amsmath,amssymb,
 aps,
 pre,
 longbibliography,
 lengthcheck,%
]{revtex4-1}

\usepackage{graphicx}
\usepackage{dcolumn}
\usepackage{bm}
\usepackage{hyperref}
\graphicspath{{Figures/}}

\begin{document}

\title{Active particles in reactive disordered media: how does adsorption affects diffusion?}

\author{R. Salgado-Garc\'{\i}a} 
\email{raulsg@uaem.mx}
\affiliation{Centro de Investigaci\'on en Ciencias-IICBA, Universidad Aut\'onoma del Estado de Morelos. Avenida Universidad 1001, Colonia Chamilpa, 62209, Cuernavaca Morelos, Mexico.} 

\date{\today}

\begin{abstract}
In this work we study analytically and numerically the transport properties of non-interacting active particles moving on a $d$-dimensional disordered media. The disorder in the space is modeled by means of a set of non-overlapping spherical obstacles. We assume that obstacles are \emph{reactive} in the sense that they react in the presence of the particles in an attractive manner: when the particle collides with an obstacle, it is attached during a random time (\emph{adsorption time}), i.e., it gets adsorbed by an obstacle; thereafter the particle  is detached from the obstacle and continues its motion in a random direction. We give an analytical formula for the effective diffusion coefficient when the mean adsorption time is finite. When the mean adsorption time is infinite, we show that the system undergoes a transition from a normal to anomalous diffusion regime. 
We also show that another transition takes place in the mean number of adsorbed particles: in the anomalous diffusion phase all the particles become adsorbed in the average. 
We show that the fraction of adsorbed particles, seen as an order parameter of the system, undergoes a second-order-like phase transition, because the fraction of adsorbed particles is not differentiable but changes continuously as a function of a parameter of the model.

\end{abstract}

\pacs{05.40.-a,05.60.-k,05.10.Gg}

\maketitle

\section{Introduction}
\label{sec:intro}

Active particles have the ability to take energy from the environment and drive themselves out of equilibrium. This characteristic endows a system of active particles with novel properties such as the emergence of collective behavior~\cite{ramaswamy2010mechanics,Marchetti2013Hydrodynamics,Bechinger2016Active}. Active particles models have been used to model several systems of real life such as microswimmers~\cite{Elgeti2015Physics}, cell tissues~\cite{Giavazzi2018Flocking}, microtubuli~\cite{Sanchez2012Active}, flocks of birds~\cite{Bialek2012Statistical} insects swarms~\cite{Sinhuber2017Phase,Attanasi2014Collective}, or shoaling fish~\cite{Herbert-Read18726,Jhawar2020Noise}, just to mention a few examples. The study of these systems has revealed several remarkable properties in both, interacting and non-interacting active particle models. For instance, on one hand the interacting active particle systems have shown to exhibit the emergence of large scale collective motion~\cite{Vicsek2012Collective,ramaswamy2010mechanics,Sevilla2016Diffusion,Chepizhko2013Optimal} as well as large-scale instabilities~\cite{Marchetti2013Hydrodynamics}. On the other hand, non-interacting active particle systems have shown to exhibit complex non-equilibrium transients in the mean square fluctuations~\cite{Peruani2007Self} and anomalous velocity distributions~\cite{Golestanian2009Anomalous} among other other interesting features~\cite{Biswas2020First,Bechinger2016Active,su2017colloidal}.

Particularly some attention has been addressed to the problem of non-interacting active particles in disordered and complex media~\cite{Bechinger2016Active,Chepizhko2013Diffusion,bertrand2018optimized,Jakuszeit2019Diffusion,Zeitz2017Active}. For instance, in Ref.~\cite{Chepizhko2013Diffusion} it was studied the transport properties of active particles in a heterogeneous  two-dimensional space in which the heterogeneity is modeled by randomly distributed obstacles. In that work, the authors reported the occurrence of spontaneous trapping of active particle at large obstacle density. In Ref.~\cite{bertrand2018optimized} it was considered a system of run-and-tumble particles, which is a prototypical model of self-propelled particles, in crowded environments. They found that this simplified version of active particle systems can be dealt with in an analytical way in $d$-dimensions, thus providing explicit formulas for the diffusivity of run-and-tumble particles.

In this work we are mainly interested in the study of the diffusion of non-interacting active particles in a disordered $d$-dimensional reactive medium. The latter means that the disordered medium interacts (in this case, attractively) with the particles. This interaction is such that the particles are adhered to the obstacles surface,  a phenomenon which is commonly referred to as \emph{adsorption}. In our model the disorder of the medium is modeled by means of a set of randomly located obstacles in a $d$-dimensional space, and the obstacles have the property of exerting a force over the particles via an ``attractive potential''. The result of this attractive potential is that the particles get attached during a time that we consider as a random variable. Once the particle is detached from the obstacle, it continues its motion in a random direction. We are interested in describing the diffusion properties of this model and the effect of the adsorption over the diffusivity of the active particles. Particularly we will show that there is a transition from normal to anomalous diffusion when the mean adsorption time goes to infinity. This transition is accompanied by a transition in which all the particles, in the average, are adsorbed, which is a kind of trapping effect for the active particles. The trapping transition we report here is induced by  different mechanism of the one reported in Ref.~\cite{Chepizhko2013Diffusion,Morin2017Diffusion} where the transition depends on the obstacle density; here the trapping transition is a consequence of the adsorption properties of the medium rather than the obstacle density. 

This paper is organized as follows. In Sec.~\ref{sec:model} we present the model of active particles, the disorder and the adsorption properties  of the medium. In Sec.~\ref{sec:normal} we compute the effective diffusion coefficient when the mean adsorption time is finite, which is the condition to have normal diffusion for the active particles. We test our formulas for the diffusion coefficient in two a three dimensions by means of numerical simulations. In Sec.~\ref{sec:anomalous} we assume that the mean adsorption time diverges, thus the active particles no longer diffuses in a normal way,  leading to a transition from normal to anomalous subdiffusion. In this case we are interested in computing the anomalous diffusion exponent, for which we obtain an analytical expression. We also analyze the fraction of adsorbed particles and we compared our predictions with numerical simulations. Finally in Sec.~\ref{sec:conclusions} we give the main results and conclusions of this work.

\section{Model}
\label{sec:model}

The model we will consider consists of an ensemble of non-interacting active particles moving on a $d$-dimensional space. The space is assumed to be non-homogeneous and the inhomogeneities are modeled by means of a set of non-overlapping hard hyper-spherical obstacles of radius $a$. We will denote by $\phi$ the \emph{packing fraction} parameter, which is the fraction of volume occupied by the obstacles. To be precise, if we consider a finite hyper-cube $\mathcal{D} \subset \mathbb{R}^d$ volume in $\mathbb{R}^d$ of length $L$, and if denote by $M $ the number of obstacles randomly placed in  $\mathcal{D}$, then the packing fraction $\phi$ is defined as
\begin{equation}
 \phi := \frac{ M V(\mathcal{S}) }{V(\mathcal{D})}
\end{equation}
where $V(\mathcal{S})$ is the volume occupied by a single obstacle in $\mathcal{D}$ and $V(\mathcal{D}) = L^d$ is the volume ob the hyper-cube $\mathcal{D}$. Since we are considering hyper-spheres as obstacles it is clear that $V(\mathcal{S}) = S_d(a)$ where $S_d(a)$ stands for the volume of a $d$-dimensional hyper-sphere of radius $a$,
\begin{equation}
S_d(a) := \frac{\pi^{d/2} a^d}{\Gamma\left(1+\frac{d}{2}\right)},
\end{equation}
being $\Gamma$ the well-known Euler Gamma function.

\begin{figure}[t]
\begin{center}
\scalebox{0.45}{\includegraphics{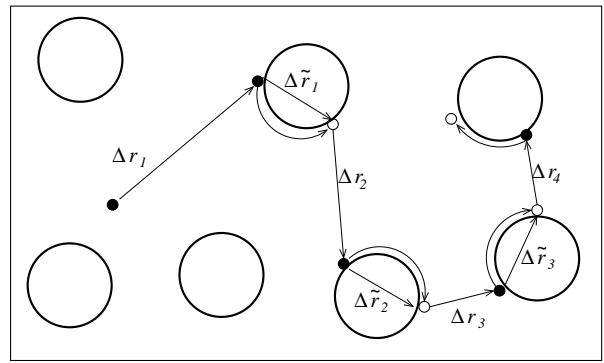}}
\end{center}
     \caption{
         Active particles in a 2D disordered medium. The disorder is introduced in the model by means of a set of hard obstacles, fixed in space having the role of scatters for the particle motion. The particle move freely outside the obstacles with constant speed $v$. The trajectory of the particle is described as a sequence of successive displacements. The (random) vectors $\Delta \mathbf{r}_1$, $\Delta \mathbf{r}_2$, $\Delta \mathbf{r}_2$, \dots are displacements of the particle between successive collisions. When the particle collides with an obstacle, it remains on the obstacle surface. After a random time the particle leaves the surface and continues its motion outside the obstacles in a straight line. The net displacements performed during the adsorption stage are denoted by  $\Delta \tilde{\mathbf{r}}_1$, $\Delta\tilde{\mathbf{r}}_2$, $\Delta \tilde{\mathbf{r}}_2$, \dots. Clearly the sum of all these displacements describe the trajectory of the particle.
              }
\label{fig:fig01}
\end{figure}

The particles are active in the sense that they move with a constant speed $v$ along a direction that might change due to the collision with the obstacles. 
When an active particle collides with an obstacle it gets adsorbed for a while. The time span during which the particles remains adsorbed will be referred to as the \emph{adsorption time}. Our model assumes that the adsorption times are drawn at random with a given distribution characterized by a probability density function (PDF) denoted by $\rho_\tau(t)$.
In the following sections we will made some assumptions about the nature of this probability distribution. 
Once the particle is desorbed, it continues its motion with the same constant speed, but in a random direction. The new direction is chosen at random with uniform distribution over all the possible directions in the $d$ dimensional space. Thereafter, the particle moves in a straight line along this new direction, with the constant speed $v$, until a new collision of the particle with an obstacle takes place. Fig.~\ref{fig:fig01} shows a schematic representation of the active particle motion in a non-homogeneous two-dimensional space.
%
%

In order to describe the transport properties of the system we will consider the motion of a single particle. This can be done because we assume that the active particles do not interact each other. Actually, a given particle just interacts with the ensemble of hard hyper-spheres (obstacles) or, in other words, the particles only interact with the disordered medium. 

Let  $\mathbf{r}(t) :=(x_1(t),x_2(t),\dots, x_d(t))  $ denote the position of the particle at time $t$ and let $\mathbf{v}(t)$ be its velocity. At time $t=0$ the particle starts its motion at $\mathbf{r}_0 :=\mathbf{r}(0) = (x_1(0),x_2(0),\dots, x_d(0))$ with velocity $\mathbf{v}(0) = v\mathbf{e}_{1}$ where $\mathbf{e}_{1}$ is a unitary vector in $\mathbb{R}^d$ drawn at random with a uniform distribution.

Once the particle starts its motion, it moves in a straight line with constant speed $v$ until a collision with an obstacle takes place. Then the particle gets \emph{adsorbed} by the obstacle surface and surrounds the obstacle during a random time  (the adsorption time) $\tau$ which is drawn at random with density $\rho_\tau(t)$.  Once the particle is detached from the obstacle, it moves again with constant speed $v$ in a new random direction $\mathbf{e}_{1}$, which is a unitary vector in $\mathbb{R}^d$ drawn at random with uniform distribution. This is the main process by which the particle spreads in space. Thereafter, the particle moves in a straight line  with constant speed until a new collision occurs. It can be observed that when the particle is not ``attached'' to the obstacles,  the magnitude $v := |\mathbf{v}(t)|$ of the vector velocity is constant, and its direction is randomly changing every time a collision occurs.  In Fig.~\ref{fig:fig01} it can be appreciated graphically the main features of the model.

At this point it is important to state that we consider all the physical quantities as dimensionless. This can be done as follows. First we choose (conveniently) a length $L_c$ and a time $T_c$. Then, if $\tilde v $ is the magnitude of the velocity of a particle and $v$ is the corresponding dimensionless velocity, then 
\begin{equation}
v :=\frac{T_c}{L_c } \tilde v.  
\end{equation}
All the lengths and times appearing in the model can also be put in its dimensionless version by dividing them by $L_c$ and $T_c$ respectively. For instance if $\tilde L$ is the length of the box $\mathcal{D}$, then, the dimensionless length of the box is defined as $L := \tilde{L}/L_c$ and if $\tilde t$ is the time, the dimensionless time $t$ is defined as $t:= \tilde t /T_c$. Notice that the choice of $L_c$ and $T_c$ is arbitrary, thus we can say that the value of $v$, in certain sense, fix the units to be used in the system.  

The main goals in this work is to study the diffusion properties as well as the adsorption properties of this model. In particular when the conditions for the occurrence of normal diffusion are guaranteed we will be interested in the effective diffusion coefficient. If we denote by $\mathbf{R}(t)$ the total displacement of the particle from its initial condition, i.e., $ \mathbf{R}(t) := \mathbf{r}(t) - \mathbf{r}(0) $, then the effective diffusion coefficient is defined as,
\begin{equation}
D_{\mathrm{eff}} := \lim_{t\to \infty} \frac{\langle  |\mathbf{R}(t)|^2 \rangle -  |\langle \mathbf{R}(t) \rangle|^2}{2\,d\,t},
\end{equation}
where the  expected value  $\langle \cdot \rangle $ is computed by averaging over an ensamble of independent trajectories $\mathbf{r}(t)$.

\section{Normal diffusion}
\label{sec:normal}


Now we shall present some calculations for obtaining an analytical expression for the effective diffusion coefficient. To this end we will made use of the central limit theorem in order to obtain the asymptotic statistical properties of the net displacement $\left| \mathbf{r}(t) - \mathbf{r}(0)\right|$ for large $t$. This procedure is similar to the one used for obtaining the transport properties (such as the diffusion coefficient and the effective particle current) in one-dimensional disordered systems~\cite{Salgado2013Normal,Salgado2014Effective,Salgado2016Normal,Hidalgo2017Scarce}. 

In this section we will assume that the distribution of adsorption time has finite mean and variance,  which makes valid the central limit theorem. Then, for the subsequent calculations it is not necessary to state the explicit form of the distribution $\rho_\tau$. Rather, it is enough to assume that 
\begin{eqnarray}
\bar{\tau} &:=& \langle \tau \rangle =\int t \rho_{\tau}(t)dt,
\\
\sigma_\tau^2 &:=& \langle \tau^2 \rangle - \langle \tau \rangle^2= \int t^2 \rho_{\tau}(t)dt - \left(\int t \rho_{\tau}(t)dt \right)^2.
\end{eqnarray}
are both finite.

\subsection{Diffusion coefficient}
\label{ssec:Deff}

The goal in this section is to describe the motion of an active particle on the disordered medium. As stated above, we consider a particle starting its motion at the point $\mathbf{r}(0) = \mathbf{r}_0
$ with an initial random direction $\mathbf{e}_1$. Then the particle moves in a straight line which can be written as
\begin{equation}
\label{eq:eq:rt0}
\mathbf{r}(t)  = \mathbf{r}_0 + v t\,\mathbf{e}_{1},
\end{equation}
where $\mathbf{e}_{1}$ is a unitary vector in $\mathbb{R}^d$ and $v$ is the constant speed of the active particle.  Eq.~\eqref{eq:eq:rt0} is the trajectory of the particle before the particle collides for the first time with an obstacle. Let $t_1$ the time at which the first collision occurs and $\ell_1 :=  |\mathbf{r}(t_1) - \mathbf{r}(0)|$ the distance traveled by the particle from the initial position  until the first collision. We will denote by $\Delta t_1 := t_1 - t_0$ (with $t_0 := 0$)  the time elapsed from the beginning of the motion until the first collision takes place. Clearly $\ell_1$ and $\Delta t_1$ are related by 
\begin{equation}
\ell_1 = v \Delta t_1.
\end{equation}
Moreover, it is also clear that the total displacement $\Delta \mathbf{r}_1 := \mathbf{r}(t_1) - \mathbf{r}(0) $ can be written as $\Delta \mathbf{r}_1 = v \mathbf{e}_{1} \Delta t_1 $.

Next, once the collision has occurred, the particle gets attached to the obstacle during a time $\tau_1$ which, as we said before, is a random variable with a PDF $\rho_\tau$. After that, the particle is detached from the obstacle and moves with a constant speed $v$ in a new independent random direction defined by $\mathbf{e}_2$. We assume that the particle leaves the obstacle surface  at some point which is different from the point at which it was adsorbed. This mimics in some way a diffusion process occurring at the surface of the obstacle, thus endowing the particle with a kind of ``scattering'' mechanism by it interaction with the obstacle. Therefore, during the adsorption stage, the particle also performs a displacement, as we illustrate in Fig.~\ref{fig:fig01}. We will denote such a displacement by $\Delta \tilde{\mathbf{r}}_1$. 

The above described process repeats again: after the particle is detached from the first obstacle, it moves in a straight line until it reaches the next obstacle, and so on. This process defines sequences of displacements $\{\Delta \mathbf{r}_j \, : \, 1\leq j\leq n \}$ and  $\{\Delta \tilde{\mathbf{r}}_j \, : \, 1\leq j\leq n \}$  which describe the trajectory of the particle up to certain time $t$. The total displacement of the particle $ \mathbf{R} (t) =\mathbf{r}(t) -\mathbf{r}(0) $ can be written as
\begin{equation}
\label{eq:displacement-sum}
 \mathbf{R}(t) = \sum_{j=0}^{n} \left( \Delta \mathbf{r}_j + \Delta \tilde{\mathbf{r}}_j \right).
\end{equation}
In Appendix~\ref{ape:1} we show that the displacement $\Delta \tilde{\mathbf{r}}_j$  is a random vector of zero mean, $\langle\Delta \tilde{\mathbf{r}}_j  \rangle  = \mathbf{0}$ and variance $\langle\Delta | \tilde{\mathbf{r}}_j  |^2 \rangle  - \big| \langle\Delta \tilde{\mathbf{r}}_j  \rangle\big|^2=  2a^2$. On the other hand, the random vector $\Delta \mathbf{r}_j$ can be written as, 
\begin{equation}
\label{eq:Drj}
\Delta \mathbf{r}_j := \ell_j \mathbf{e}_{j}, \qquad \mbox{for }1\leq j \leq n,
\end{equation}
where $\ell_j$ is the distance traveled by the particle from the $(j-1)$th to the $j$th collision. Such a distance turns out to be a random variable whose distribution is known as the \emph{free path distribution}~\cite{Bourgain1998Distribution}. It is known that the free path distribution $f(x)$ of a particles moving linearly in a $d$-dimensional space is approximately distributed according to an exponential distribution~\cite{Adib2008Random,Binglin1993Chord}, whose PDF can be expressed as
\begin{equation}
\label{eq:pdf-free-path}
f(x) = \left\{ \begin{array} 
            {r@{\quad \mbox{ if } \quad}l} 
\frac{1}{\lambda} e^{-x/\lambda}  &  x >0  \\ 
0    & x<0,
             \end{array} \right.
\end{equation}
where $\lambda$ is known as the \emph{mean free path} and can be written in terms of the parameters of the obstacle distribution as follows~\cite{Adib2008Random,Binglin1993Chord}
\begin{equation}
\label{eq:mean-free-path}
\lambda =  a \frac{s_{d}}{s_{d-1}} \frac{1-\phi}{\phi},
\end{equation}
where $s_d$ is defined as the volume of a hyper-sphere of radius one, i.e., $s_d := S_d(1)$.

It is clear that the free paths $\ell_j$ between collisions are independent because we assume that $i$) the obstacle distribution are uniformly and \emph{independently} distributed and $ii$) the random directions randomly drawn after the adsorption time are independent each other. These two assumptions imply that no correlations are present among successive free paths. In other words, we can say that the random variables $\{\ell_j \, : \, 1\leq j\leq n \}$ are independent and identically distributed with the PDF $f(x)$. Therefore, the displacements $\{ \Delta \mathbf{r}_j  \, : \, 1\leq j\leq n \}$ are independent and identically distributed random vectors. The latter is a consequence of the assumption that the random directions, along which the particle continues its motion after a collision, are also independent.

On the other hand, the total time $t$ necessary for the particle to perform the net displacement  $ \mathbf{R} (t)$ can be written as,
\begin{equation}
\label{eq:time-disp-sum}
t = \sum_{j=1}^n \Delta t_j + \sum_{j=1}^n \tau_j,
\end{equation}
which, as we see, relates the variable $n$ (the number of collisions) and the time $t$. 
By definition we have that $\{\tau_j \, : \, 1\leq j\leq n \}$  are independent and identically distributed random variables. As we saw above, the random variable $\Delta t_j$ is related to the random variable $\ell_j$ by a linear transformation, 
\begin{equation}
\ell_j = v \Delta t_j, 
\end{equation}
which makes $\Delta t_j$ to have, up to a constant factor, the same statistical properties of $\ell_j$ which we described above. To be precise, $\Delta t_j$ has exponential distribution with a parameter $\theta := v/\lambda$, i.e., the PDF of $\Delta t_j$ is given by
\begin{equation}
g(t) = \left\{ \begin{array} 
            {r@{\quad \mbox{ if } \quad}l} 
\frac{1}{\theta} e^{-t/\theta}  &  t >0  \\ 
0    & t<0.
             \end{array} \right.
\end{equation}

The next step for obtaining the asymptotic behavior of the trajectory $\mathbf{r}(t)$ is to apply the limit theorems to the sum of random variables appearing in Eqs.~(\ref{eq:displacement-sum}) and~(\ref{eq:time-disp-sum}). This technique has been proved to be useful for determining the exact diffusion coefficients in one-dimensional disordered media~\cite{Salgado2013Normal,Salgado2014Effective,Salgado2016Normal,Hidalgo2017Scarce}.

Now we proceed to compute the effective diffusion coefficient. The first step consists in applying the central limit theorem to the sum of random vectors appearing in Eq.~(\ref{eq:displacement-sum}) and the sum of random variables in Eq.~(\ref{eq:time-disp-sum}). 

The central limit theorem~\cite{gnedenko1954limit,meerschaert2001limit} states that the sum of random vectors $\mathbf{S}_n$,  defined as,
\begin{equation}
\mathbf{S}_n := \sum_{j=1}^{n} \left( \Delta \mathbf{r}_j + \Delta \tilde{\mathbf{r}}_j \right),
\end{equation}
converges to a normal distribution in the limit of $n\to\infty$ if it is appropriately normalized. To be precise,
\begin{equation}
\label{eq:CLT-SD}
\lim_{n\to \infty} \frac{\mathbf{S}_n - \langle \mathbf{S}_n \rangle  }{\sqrt{n}} =
\lim_{n\to \infty} \frac{\sum_{j=1}^{n} \left( \Delta \mathbf{r}_j - \Delta \tilde{ \mathbf{r}}_j \right)    }{\sqrt{n}} = \mathbf{Z},
\end{equation} 
where $\mathbf{Z}$ is a random vector with multivariate normal distribution of zero mean, $\langle \mathbf{Z}\rangle = \mathbf{0}$ whose variance matrix  is denoted by $  \mathbf{\Sigma}$ and is given in Appendix~\ref{ape:1}. This result allows to approximate the finite sum of displacements as,
\begin{equation}
\label{eq:sumD-approx}
\mathbf{R}(t)  = \sum_{j=0}^{n}\left( \Delta \mathbf{r}_j - \Delta \tilde{ \mathbf{r}}_j \right) \approx \sqrt{n} \mathbf{Z}, \qquad \mbox{for} \quad n\gg 1.
\end{equation}
Notice that  $n$ is the number of collisions occurred up to time $t$, thus, if $n\to \infty$ then $t\to \infty$.

Next, we will use the same technique in order to approximate the sum of random times,
\begin{equation}
\label{eq:sum-int}
\sum_{j=1}^n \left( \Delta t_j +\tau_j \right). 
\end{equation}
In order to apply the central limit theorem, we denote by $\mu $ the mean of the random variable $ t_j +\tau_j$ and by $\sigma$ its variance, i.e.,
\begin{eqnarray}
\mu &:=&\langle  \Delta t_j  + \tau_j \rangle = \langle  t_j \rangle + \langle\tau_j \rangle,
\\
\sigma^2 &:=& \langle (\Delta t_j +\tau_j)^2 \rangle - \mu^2.
\end{eqnarray}
Some direct calculations show that we can write $\mu$ and $\sigma^2$ as,
\begin{eqnarray}
\label{eq:mu-sig-explicit}
\mu &=& \frac{\lambda}{v} + \bar{\tau},
\\
\sigma^2 &=& \frac{\lambda^2 }{v} + \sigma_\tau^2.
\end{eqnarray}
Therefore, the sum of random times en Eq.~\eqref{eq:sum-int} comply with the central limit theorem as follows
\begin{equation}
\lim_{n\to \infty}\frac{\sum_{j=1}^n \left(\Delta  t_j +\tau_j \right) - n \mu }{\sqrt{n}\sigma} =  Z_0,
\end{equation}
where $Z_0$ is a random variable with standard normal distribution, i.e., a random variable with normal distribution with zero mean, $\langle Z_0 \rangle =0$ and variance $\langle Z_0^2 \rangle -  \langle Z_0 \rangle^2 =1$ .

Then, the finite sum of times can be approximated by
\begin{equation}
\label{eq:sumT-approx}
\sum_{j=1}^n \left( \Delta t_j +\tau_j \right) \approx  n \mu + \sqrt{n}\sigma  Z_0,    \qquad \mbox{for} \quad n\gg 1.
\end{equation}

It is important to remark that Eqs.~(\ref{eq:sumD-approx}) and~(\ref{eq:sumT-approx}) allow us  to obtain the  statistical properties of the net displacement $ \mathbf{R}(t)$ for large $t$. First notice that, as expressed in Eq.~(\ref{eq:displacement-sum}), $ \mathbf{R}(t)$ is not explicitly a function of time $t$. Actually the information of the total time $t$ necessary for the particle to perform the net displacement  $ \mathbf{R}(t)$ is contained in Eq.~(\ref{eq:time-disp-sum}). Then, the strategy followed here is to  use the approximation~(\ref{eq:sumT-approx}) for obtaining an expression for the number of collisions $n$ as a function of the total time $t$. Hence, this result is used  within~(\ref{eq:sumD-approx}) in order to have the total displacement $ \mathbf{R}(t)$ explicitly in terms of the time $t$. Notice also that, the larger time $t$, the larger number of collisions $n$. Therefore, expressions~(\ref{eq:sumD-approx}) and~(\ref{eq:sumT-approx}), which are valid for large $n$, are also valid for large time $t$.

Within this approximation we have that the total time $t$ as a function of the number of collisions $t$ is given by
\begin{equation}
\label{eq:for-t-clt}
n \mu + \sqrt{n}\sigma  Z_0 \approx t, \quad \mbox{for large t}.
\end{equation}
It is clear that, if the time $t$ is fixed, then the number of collisions performed by the particle is a random variable. The above equations relates these variables in the limit of large $t$ by means of the central limit theorem.  According to  Ref.~\cite{Salgado2013Normal}, Eq.~(\ref{eq:for-t-clt}) can be solved for $n$ giving  
\begin{equation}
\label{eq:t-clt}
n\approx \frac{t}{\mu} - \frac{\sigma\, t^{1/2}}{\mu^{3/2}} Z_0.
\end{equation}

The next step consists in substituting $n$ given in the above expression, into the asymptotic expression for the net displacement~(\ref{eq:sumD-approx}), which results in,
\begin{equation}
\label{eq:Dr(t)-asympt-0}
\mathbf{R}(t) \approx n^{1/2} \mathbf{Z} = \left( \frac{t}{\mu} - \frac{\sigma\, t^{1/2}}{\mu^{3/2}} Z_0 \right)^{1/2} \mathbf{Z}.
\end{equation}
We can see that, the leading order as $t\to \infty $ in the above expression is $t^{1/2}$, thus obtaining the following asymptotic expression for the net displacement,
\begin{equation}
\label{eq:l-asympt}
 \mathbf{R}(t)  \approx \left( \frac{t}{\mu} \right)^{1/2}  \mathbf{Z},
\end{equation}
for $t\to \infty$. 

From Eq.~\eqref{eq:l-asympt} we see that $\langle \mathbf{R}(t)  \rangle =0$, which allows us to write the diffusion coefficient as
\begin{eqnarray}
D_{\mathrm{eff}} &=& \lim_{t\to \infty } \frac{ \langle |\mathbf{R}(t)|^2 \rangle }{2\,d\,t}
\nonumber
\\
&=& \lim_{t\to \infty } \frac{  \langle | t^{1/2}\mu^{-1/2}   \mathbf{Z} |^2 \rangle    }{2\,d\,t} =  \frac{  \langle |  \mathbf{Z} |^2 \rangle    }{2\,d\, \mu}. 
\end{eqnarray}

In Appendix~\ref{ape:1} we prove that, 
\begin{equation}
\label{eq:Z2}
\langle |  \mathbf{Z} |^2 \rangle  = 2\lambda^2 + 2 a^2.
\end{equation}
which leads to the following expression for the diffusion coefficient, 
\begin{eqnarray}
D_{\mathrm{eff}} &=& \frac{\lambda^2  + a^2 }{d \mu }.
\end{eqnarray}
Recalling Eq.~(\ref{eq:mu-sig-explicit}), the definition of $\mu$, the above expression for $D_{\mathrm{eff}}$ can be rewritten as,
\begin{eqnarray}
\label{eq:Deff-final}
D_{\mathrm{eff}} &=&  \frac{1}{d}\frac{\lambda^2 + a^2}{ \bar{\tau} + \lambda/v  },
\end{eqnarray}
where $\lambda$ is the mean free path given in Eq.~(\ref{eq:mean-free-path}).

\subsection{Numerical experiments}

%
\begin{figure}[t]
\begin{center}
\scalebox{0.36}{\includegraphics{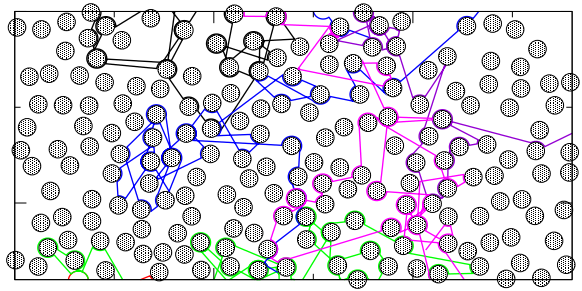}}
\end{center}
     \caption{ Simulation of active particles diffusing in a two-dimensional disordered medium. The particles move in a straight line at constant speed outside the obstacles. When a particle collides with an obstacle it gets adsorbed during a random time interval. Once the particles gets desorbed, they continue their motion in random directions. For this simulation the packing fraction is $\phi = 0.3927$. 
                   }
\label{fig:fig02}
\end{figure}
%
In order to test our formula for the diffusion coefficient we perform numerical simulations of the system of active particles in two and three dimensions. For the two-dimensional case we consider a system of $N = 10^3$ particles and $M = 20\times 10^4$ obstacles which are placed randomly with uniform distribution in a  two-dimensional square box $ \mathcal{D}$ of wide $L=10^{4}$. We set the particle velocity to $v=1$ and the distribution of the adsorption times is chosen as an exponential distribution with mean $\bar{\tau} = 1$. We use periodic boundary conditions for the particle dynamics in $ \mathcal{D}$. Notice that the volume of $ \mathcal{D}$ is $ V(\mathcal{D}) = 10^8$. We choose several values for the radius of the obstacles $a$, ranging from $a = 2.5$ to $a = 50$. These values result in several values for the packing fraction $\phi$.

In Fig.~\ref{fig:fig02} we show some examples of trajectories of active particles with different initial conditions. We use a packing fraction $\phi \approx 0.3927$ corresponding to $a = 25$.  We see that the time that the particles remains ``adhered'' to the obstacle surface during a random time, before the particles continues its motion in an arbitrary direction. We interpret this characteristic of our model as if the particle were performing a surface diffusion while it is adsorbed. The latter can be though as similar, for instance, to the mechanism of facilitated diffusion in DNA~\cite{Lomholt8204,Mirny_2009}, in which the involved particles might execute diffusion in different dimensions. 
%
\begin{figure}[t]
\begin{center}
\scalebox{0.3}{\includegraphics{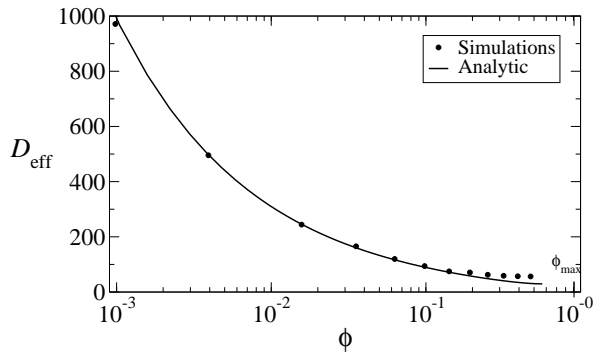}}
\end{center}
     \caption{Effective diffusion coefficient for active particles in two dimensions. The effective diffusion coefficient estimated from numerical experiments are obtained for several values of packing fraction $\phi$, which are shown in filled circles. The approximated formula given in Eq.~\eqref{eq:Deff-final} is represented by the solid line. We observe that the numerical experiments and the theoretically computed diffusion coefficient are consistent within the accuracy of our numerical simulations. The value $\phi_{\mathrm{max}}$ corresponds to the maximal packing fraction which is nearly $\phi \approx 0.907$~\cite{torquato2013random}.
              }
\label{fig:fig03}
\end{figure}
%

Next we perform the simulation of the $N = 10^3$ particles obtaining a the same number of trajectories. The total simulation time is set to $T = 5\times 10^5$. Once we obtain all the trajectories we estimate the mean square displacement averaging over all the simulated trajectories which allows to estimate the effective diffusion coefficient. In Fig.~\ref{fig:fig03} we show the effective diffusion coefficient for several values of the packing fraction, which corresponds to the radii chosen for the simulations: $a =  2.5\, m $  for $m=1,2,\dots, 10$, corresponding to the filled circles shown in Fig.~\ref{fig:fig03}. In the same figure we show our analytical formula (solid black line)  given in Eq.~\eqref{eq:Deff-final}.

\begin{figure}[t]
\begin{center}
\scalebox{0.3}{\includegraphics{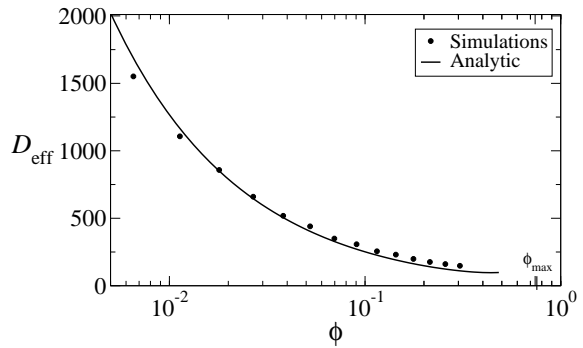}}
\end{center}
     \caption{
     Effective diffusion coefficient for active particles in three dimensions. The effective diffusion coefficient estimated from numerical experiments are obtained for several values of packing fraction $\phi$, which are shown in filled circles. The approximated formula given in Eq.~\eqref{eq:Deff-final} is represented by the solid line. We observe that the numerical experiments and the theoretically computed diffusion coefficient are consistent within the accuracy of our numerical simulations. The value $\phi_{\mathrm{max}}$ corresponds to the maximal packing fraction which is nearly $\phi \approx 0.740$~\cite{torquato2013random}.
              }
\label{fig:fig04}
\end{figure}

For the three dimensional case we consider a system of $N = 10^3$ particles and $M = 20\times 10^4$ obstacles which are placed randomly with uniform distribution in a  three dimensional cube $ \mathcal{D}$ of wide $L=10^{4}$. We set the particles velocity to $v=1$ and the adsorption times are distributed according to an exponential distribution with mean $\bar{\tau} = 1$ . We use periodic boundary conditions for the particle dynamics in $ \mathcal{D}$.  We choose several values for the radius of the obstacles $a$, ranging from $a = 40$ to $a = 180$. These values fix the packing fraction.  The total simulation time is set to $T = 5\times 10^6$. Once we obtain all the trajectories, we estimate the mean square displacement averaging over all the trajectories which allows to estimate the effective diffusion coefficient. In Fig.~\ref{fig:fig04} we show the effective diffusion coefficient for several values of the packing fraction corresponding to the radii chosen for the simulations: $a = 40 + 5\, m $  for $m=0,1,2,\dots, 14$, corresponding to the filled circles shown in Fig.~\ref{fig:fig04}. In the same figure, solid line represents the  analytical formula given in Eq.~\eqref{eq:Deff-final}.
%
%

We can observe that in both cases the analytical formula predicts satisfactorily the behavior of the diffusion coefficient. As we see, our formula seems to be less satisfactory as we increase the packing fraction. This phenomenon has been also observed in the study of the effective diffusion in disordered Lorentz gas. This loss of accuracy might have its origin in the fact that the free path distribution given in Eq.~\eqref{eq:pdf-free-path} is actually an approximation (valid for $\phi \ll 1$) to the real free path distribution~\cite{Adib2008Random,Binglin1993Chord}. This, together with the statistical errors in numerical simulations might cause this effect and requiere a further detailed study. Nevertheless, these numerical simulations shows that the analytical approach presented here is satisfactory enough to analyze other cases that deserve our attention.

\section{Anomalous diffusion}
\label{sec:anomalous}

In this section we explore the case in which the mean adsorption time is no longer finite. As we mentioned above, in this case the system no longer exhibits normal diffusion and instead, the square displacement grows, in the average, lower than linear in time,
\begin{equation}
\label{eq:asymp-DR}
\langle|  \mathbf{R}|^2 \rangle - \big| \langle  \mathbf{R} \rangle \big|^2  \propto t^{\beta},
\end{equation}
with $0< \beta <1$. In the following we will refer to $\beta$ as the \emph{anomalous diffusion exponent}.

It is known that a sum of random variables having no finite variance nor finite mean, does not fulfill the central limit theorem. Nevertheless, such a sum of independent random variables still converges (appropriately normalized) to a stable law~\cite{gnedenko1954limit,meerschaert2001limit}. 
When the central limit theorem is broken in this class of systems, it has been show the occurrence of a transition from normal to anomalous diffusion~\cite{Salgado2013Normal,Salgado2014Effective,Salgado2016Normal,Hidalgo2017Scarce} and the limit theorems to stable laws allow the computation of the anomalous diffusion exponent $\beta$ in Eq.~\eqref{eq:asymp-DR}.

\subsection{Anomalous exponent}

When we deal with the case in which the mean adsorption time is no longer finite. In order to analyze such a case, we introduce the model for the distribution of adsorption times, which consists in choosing the PDF $\rho_\tau  $ as a power-law,
\begin{equation}
\label{eq:tau-powlaw}
\rho_\tau(t) = \left\{ \begin{array} 
            {r@{\quad \mbox{ if } \quad}l} 
\alpha t^{-\alpha -1}    &  t \geq 1  \\ 
0    & t<1.
             \end{array} \right.
\end{equation}
Notice that the mean adsorption time diverges if $\alpha \leq 1$, otherwise the mean adsorption time is given by
\begin{equation}
\label{eq:bartau}
\bar{\tau} := \int_{-\infty}^\infty t \rho(t) dt = \int_1^\infty \alpha t^{-\alpha} dt = \frac{\alpha}{\alpha -1}.
\end{equation}
The variance of $\tau$ can be computed in a similar way, resulting in
\begin{equation}
\sigma_\tau^2 = \frac{\alpha}{\alpha -2} -\left(\frac{\alpha}{\alpha -1}\right)^2,
\end{equation}
which is clearly finite only if $\alpha > 2$.

In Appendix~\ref{ape:2} we explore the cases $i$) $0 < \alpha < 1$, $ii$) $1 < \alpha < 2$, $iii$) $\alpha =1$
 and $iv $) $\alpha = 2$ separately. The case $\alpha > 2$ is not considered here since the central limit theorem is valid for these values of $\alpha $, thus leading to normal diffusion as it was stated in Sec.~\ref{sec:normal}. The reason to treat all these cases separately is due to the fact that the normalization of the sum of random variables is different in every case, and converges to different limit distributions. However, the diffusion properties only depends on the behavior of the mean adsorption time leading to only two different regimes. Here we only summarize the results for the behavior of square displacement which, according to Appendix~\ref{ape:2}, is given by
\begin{equation}
\label{eq:beta-alpha}
 \langle|  \mathbf{R}|^2 \rangle - \big| \langle  \mathbf{R} \rangle \big|^2 \propto \left\{ \begin{array} 
            {r@{\quad \mbox{ if } \quad}l} 
t^\alpha  &   0 < \alpha \leq 1  \\ 
\frac{t}{\ln(t)}  &  \alpha = 1  \\ 
t          & \alpha > 1.
             \end{array} \right.
\end{equation}
This result shows that the system undergoes a transition from normal ($\alpha > 1$) to anomalous ($\alpha \leq 1$) diffusion. The anomalous diffusion phase is characterized by the behavior of the mean square displacement which grows in time lower that linear. For $0 < \alpha <1$ this growth is a power of $t$ with exponent $\beta = \alpha$. For $\alpha = 1$ the grow in time of the mean square displacement is still lower than linear but not a power law. Indeed, in this case such a grow is proportional to $t$ but screened by a logarithm, behavior which is known as \emph{marginal subdiffusion}. Interestingly, the case $1 <\alpha \leq 2$ leads to a normal diffusive behavior. This fact is interesting because for these values of $\alpha$ the random adsorption times have no finite variance, i.e., even having a heavy-tailed distribution for the random adsorption times with infinite variance, the system yet exhibit normal diffusive behavior. There is a plethora of examples in the literature which infinite variance (of the random variables involved in the underlying dynamics of the system) is enough for the system to exhibit anomalous behavior~\cite{BOUCHAUD1990127,METZLER20001,Salgado2016Normal,Hidalgo2017Scarce}.



%
%

\subsection{Mean number of adsorbed particles}

In this section we will compute the asymptotic behavior of the mean number of adsorbed particles in the case of adsorption times distributed according to a power law with the distribution given in Eq.~\eqref{eq:asymp-DR}. Consider an ensemble of $N$ non-interacting active particles moving in the disordered medium. Let $\mathbf{r}_k(t)$ be the trajectory of the $k$th particle for $1\leq k \leq N$. Let $S_k(t)$ denote the ``state'' of the particle, i.e., $S_k(t)=1$  if the particle is adsorbed and  $S_k(t)=0$  if not.  If we denote by $N_{\mathrm{ads}}(t)$ the number of adsorbed particles at time $t$ then it is clear that,
\begin{equation}
N_{\mathrm{ads}} (t) = \sum_{k=1}^N S_{k}(t).
\end{equation}
Then, the fraction $\gamma $ of adsorbed particles is given by
\begin{equation}
\gamma_N := \frac{ N_{\mathrm{ads}} (t) }{N} = \frac{1}{N}\sum_{k=1}^N S_{k}(t).
\end{equation}
By the law of large numbers~\cite{gnedenko1954limit} we have that, for $N \to\infty$,  the last sum converge to the mean of $S_k(t)$, which is a consequence of the fact that the $\{S_k(t)\}$ is a collection of independent random variables for a fixed $t$. Explicitly we have that  the fraction $\gamma$ of adsorbed particles for $N \to\infty$,
\begin{equation}
\gamma := \lim_{N\to\infty } \gamma_N = \lim_{N\to\infty } \frac{1}{N}\sum_{k=1}^N S_{k}(t),
\end{equation}
can be written as
\begin{equation}
\gamma =\langle S_{k}(t)\rangle.
\end{equation}
The expected value in the above expression is taken with respect to the ensemble in the thermodynamic limit. If we further assume that the process $\{ S_k(t) \, : \, t >0 \}$ is a stationary (uniquely) ergodic process we would have that $ \gamma $ does not depend on time and the ensemble average can be written as a time-average, 
\begin{equation}
\gamma = \langle S_{k}(t)\rangle = \lim_{t\to \infty} \frac{1}{t} \int_{0}^t S_k(t^\prime)dt^\prime.
\end{equation}
The last expression allows us to write the fraction of adsorbed particles in terms of the sums of random adsorption times,
\begin{equation}
\label{eq:gamma_taus}
\gamma = \lim_{t\to \infty} \frac{1}{t} \sum_{j=1}^n \tau_j(t).
\end{equation}
In Appendix~\ref{ape:3} we show that for the the model of random adsorption times ruled by a power law distribution, the fraction of the adsorbed particles in the thermodynamic-like limit is given by,
\begin{equation}
\label{eq:gamma-alpha}
\gamma = \left\{ \begin{array} 
            {r@{\quad \mbox{ if } \quad}l} 
1  &   0 < \alpha \leq 1  \\ 
\frac{\bar{\tau}}{\bar{\tau} + \frac{\lambda }{v} }    & \alpha > 1.
             \end{array} \right.
\end{equation}
Notice that the above expression is a continuous  function of $\alpha$. This can be seen as follows: notice that $ \bar{\tau} = \bar{\tau}(\alpha) $ (cf. Eq.~\eqref{eq:bartau}) and that $\tau \to \infty$ as $\alpha\to 1 $. Thus,
\begin{equation}
\lim_{\alpha \to 1} \frac{\bar{\tau}}{\bar{\tau} + \frac{\lambda }{v} }  = 1,
\end{equation}
which shows that $\gamma (\alpha)$ is continuous at $\alpha = 1$. This fact is important because, despite the fraction of adsorbed particles is continuous, it is not differentiable. This means that the system undergoes as kind of second-order phase transition manifested in the order parameter  $\gamma$.

\subsection{Numerical experiments}

In this section we test our formulas for the anomalous exponent $\beta$ and the fraction of adsorbed particles.  For this purpose we simulated the system of active particles in two dimensions. In this case we considered a system of $N = 10^3$ particles and $M = 20\times 10^4$ obstacles which were placed randomly with uniform distribution in a  two-dimensional square box $ \mathcal{D}$ of wide $L=10^{4}$. We used periodic boundary conditions for the dynamics in $ \mathcal{D}$ and set the particle velocity to $v=1$. Notice that the volume of $ \mathcal{D}$ is $ V(\mathcal{D}) = 10^8$, which, together with the choice of radius of the obstacles $a=10$, fix the packing fraction $\phi$, whose value is approximately $\phi \approx 0.01571$. For the random adsorption times we used the distribution given in Eq.~\eqref{eq:tau-powlaw} and perform the simulations for several values of $\alpha$. 

Next our goal is to characterize, from the numerically simulated trajectories, the nature of the diffusivity. To this end we estimated the exponent $\beta$ as follows. First, from the simulated trajectories we obtain the time-dependent mean square displacement $ \langle |  \mathbf{R}|^2 \rangle - \big |\langle \mathbf{R} \rangle\big|^2$ which, according to our hypothesis, it behaves in time as a power law $\propto t^\beta$. Once we we obtain the mean square displacement we made use  the least squares method to estimate $\beta$.
%
\begin{figure}[t]
\begin{center}
\scalebox{0.35}{\includegraphics{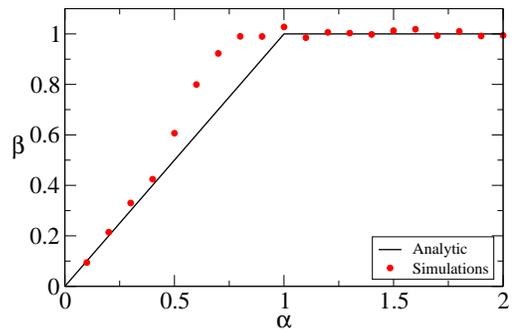}}
\end{center}
     \caption{Exponent of the mean square displacement. We show the behavior of the exponent $\beta $ of the mean square displacement as a function of $\alpha$, the parameter of the distribution of the adsorption times. The solid black line represents the theoretically obtained exponent $\beta(\alpha)$ given in Eq.~\eqref{eq:beta-alpha}. The red filled circles represent the estimated exponent $\beta$ obtained from the numerical simulations. We can see that the system undergoes a kind of second order transition, which in this case is a transition from normal ($\alpha >1$) to anomalous diffusion ($\alpha\leq 1$).
                   }
\label{fig:fig05}
\end{figure}
%

In Fig.~\ref{fig:fig05} we can see the estimated $\beta$ for several values of $\alpha$ (red filed circles). In the same figure we also display the theoretically computed exponent $\beta$ as a function of $\alpha$. We can appreciate that the numerically estimated exponent is consistent with its analytical counterpart at least for values of $\alpha$ below $\alpha = 0.5$ and above $\alpha= 1$. The deviations we observe for intermediate values of $\alpha$ can actually be explained as an effect of the slow convergence, which is in turn a consequence of the power law behavior. Below we will perform a simple analysis to estimate how slow is the convergence in this system.

Now we turn to the analysis of the mean number of adsorbed particles. From the simulated trajectories we estimated the mean fraction of adsorbed particles $\gamma$ for large times. In Fig.~\ref{fig:fig05} we show the result of the estimated $\gamma$ from the numerical simulations (red filled circles) as well as the analytically computed (solid black line) fraction of adsorbed particles $\gamma(\alpha)$ for several values of $\alpha$. As in the case of the exponent $\beta(\alpha)$ we can see that the numerical simulations are consistent with the theoretical predictions, specially  for values of $\alpha$ below $\alpha = 0.5$ and above $\alpha = 1$. The underestimation of the mean fraction of adsorbed particles can be explained as a finite-time effect, i.e., due to a slow convergence of the averaged quantities. First consider a fixed value of $\alpha$ and recall that, according to Eq.~\eqref{eq:gamma_taus}, $\gamma$ can be obtained as a limit as follows,
\begin{equation}
\gamma = \lim_{t\to \infty} \gamma_t= \lim_{t\to \infty} \frac{1}{t} \sum_{j=1}^n \tau_j(t).
\end{equation}
%
\begin{figure}[t]
\begin{center}
\scalebox{0.35}{\includegraphics{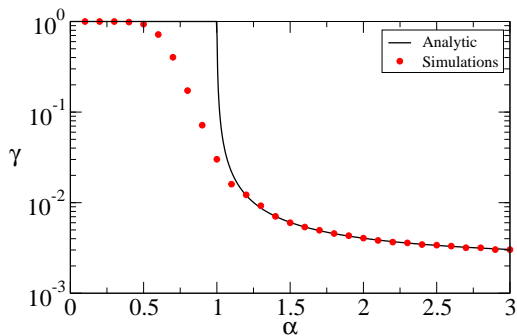}}
\end{center}
     \caption{ Fraction of adsorbed particles. We show the fraction of adsorbed particles $\gamma$ as a function of $\alpha$ obtained from numerical simulations (red filled circles) and computed from the analytical formula~\eqref{eq:gamma-alpha} (black solid line). We see that the numerically estimated $\gamma$ is consistent with the theoretical prediction. Notice that the numerical estimations are less satisfactory for $\alpha$ in the interval $(0.5,1)$, an effect that occurs due to the slow convergence of $\gamma$ in the limit $t\to \infty$.  
                        }
\label{fig:fig06}
\end{figure}
%
According to Appendix~\ref{ape:3}, if we assume that $0< \alpha < 1$ we have that $\gamma_t$ can be written alternatively  as,
\begin{equation}
\label{eq:int-gamma}
\gamma_t \approx  \frac{  W n^{1/\alpha} }{  \left( \frac{\lambda}{v} \right) n +  \left( \frac{\lambda}{v} \right) n^{1/2} Z_0 + W n^{1/\alpha} },
\end{equation}
where $n = n(t)$ is the number of collisions at time $t$. It is not hard to see that, as $t\to \infty$ (and therefore as $n\to \infty$) we have that $\gamma_t $ goes as,
\begin{eqnarray}
 \gamma_t \approx 1 - \frac{\lambda }{v W} n^{1-1/\alpha}.
\end{eqnarray}
In Appendix~\ref{ape:3} we also showed that $ n(t) \approx t^{\alpha} /W^{\alpha}$. Substituting this identity into the above expression we can obtain $\gamma_t$ explicitly as a function of $t$. Then we have
\begin{eqnarray}
\label{eq:gamma_t}
\gamma_t \approx 1 -  \frac{\lambda }{v W} \left(\frac{t^\alpha}{W^\alpha} \right)^{1-1/\alpha} \approx 1 - \frac{\lambda }{v W^\alpha} t^{\alpha-1},
\end{eqnarray}
for $0<  \alpha < 1$ and large $t$. 

Expression~\eqref{eq:gamma_t} shows that as $t$ goes to infinity, the fraction of adsorbed particles attains the value $\gamma = 1$. However, as we can also appreciate in Eq.~\eqref{eq:gamma_t} the rate of convergence is extremely slow  for $\alpha$ near (and below) $\alpha = 1$ since it goes as $t^{\alpha -1}$. Certainly, this behavior explains why the estimations obtained  from numerical simulations are less accurate for $0.5 < \alpha < 1$ than the ones made outside this interval. 
%
%
%
%

\bigskip 
\section{Conclusions}
\label{sec:conclusions}

In work we have studied a system of non-interacting active particles in a $d$-dimensional disordered  reactive medium. The disorder has been modeled as a set of non-overlapping obstacles which limits the movement of the particles throughout a hard core interaction. That the medium be reactive means that the particles interact with the medium in an attractive manner: once a particle collides with an obstacle, it gets adsorbed to it during a time which we assume to be randomly drawn with certain distribution. Next we were interested in studying the diffusion as well as the adsorption properties of this system. In particular we showed that the diffusion coefficient is finite if the distribution of the adsorption times has finite mean; actually we provided an analytical formula for the diffusion coefficient which satisfactorily fits the diffusion coefficient obtained by means of numerical simulations in two a three dimensions.  Thereafter we analyzed the behavior of the square displacement when the mean adsorption is no longer finite. We showed that in this case the system undergoes a transition from normal to anomalous subdiffusion and we were able to obtain an analytical expression for the anomalous diffusion exponent. The occurrence of this transition from  normal to anomalous diffusion is also accompanied by a transition in the fraction of adsorbed particles. When the mean adsorption time is finite  the fraction of adsorbed particles is strictly less than one. However, when the distribution of adsorption times passes to a regime in which the mean adsorption time diverges, the fraction of adsorbed particles equals one. The latter means that in such a case all the particles become adsorbed by the obstacles in the average, thus collapsing all in the surface of  the obstacles. This phenomenon in interesting because this transition can also seen at the level of distributions. In the phase in which $\gamma  < 1$, the motion of the particles is not restricted, they move in a $d$-dimensional space outside the obstacles. However, when the system reaches the phase $\gamma = 1$, then all the particles collapse their motion into a region of lower dimension. This means that the distribution of the particles passes from being absolutely continuous to singular in $\mathbb{R}^d$ with respect to Lebesgue. This phenomenon, which manifests as a change in the nature of the distribution, has shown to occur in other class of systems within the  context of thermodynamical formalism,  and such a transition is commonly referred to as \emph{freezing phase transition}~\cite{Bruin2013Renormalization}. Particularly, in Ref.~\cite{Maldonado2019Freezing} the freezing phase transition has been related to the occurrence of anomalous diffusion. Here in this work, through the analysis of the fraction of adsorbed particles we show indirectly that this might be the case in a system of active particles. Certainly, whether or not the distribution passes from continuous to singular requieres a much more mathematically rigorous analysis.

\begin{acknowledgments}
The author CONACyT by its financial support through grants A1-S-15528 and FORDECYT-PRONACES/1327701/2020.
\end{acknowledgments}

\appendix

\section{Some statistical features of random vectors}
\label{ape:1}

Let us start by characterizing the random vector displacement $\Delta \tilde {\mathbf{r}}$. As we saw in Sec.~\ref{sec:model}  $\Delta \tilde {\mathbf{r}}$ is the net displacement of the particle from the arriving point and the departing point on the obstacle surface when the adsorption takes place. The obstacle is a circle of radius $a$ and therefore the vector displacement can be written as
\begin{equation}
\Delta \tilde {\mathbf{r}} = a \left(\mathbf{e}^\prime - \mathbf{e}\right),
\end{equation}
where $a \mathbf{e}$ and $a\mathbf{e}^\prime$ are the arriving and departing points (respectively) on the obstacle surface with respect to a reference frame whose origin is the center of the obstacle. The vectors $\mathbf{e}$ and $\mathbf{e}^\prime$ are random unitary vectors. These vector are independent and uniformly distributed on all the possible directions in $d$ dimensions. The latter implies that both, $\mathbf{e}$ and $\mathbf{e}^\prime$, have zero mean, i.e.,
\begin{equation}
\langle \mathbf{e}\rangle = \langle \mathbf{e}^\prime \rangle = \mathbf{0}.
\end{equation}
We are interested in computing the expected value of $ |\Delta \tilde {\mathbf{r}}|^2 $. Notice that
\begin{equation}
|\Delta  \tilde {\mathbf{r}}|^2 = |\mathbf{e}^\prime - \mathbf{e}|^2 = 2 + \mathbf{e}.\mathbf{e}^\prime.
\end{equation}
Then we have,
\begin{equation}
\label{eq:sg_tilder}
\langle |a\Delta \tilde {\mathbf{r}}|^2 \rangle = 2a + a\langle \mathbf{e}\cdot \mathbf{e}^\prime \rangle = 2a,
\end{equation}
a result that is a consequence of the fact that $\mathbf{e}$ and $\mathbf{e}^\prime$  are independent and have zero mean.

Next we will compute the variance matrix $\mathbf{\Sigma}$ of the random vector $\mathbf{Z}$ appearing in Eq.~\eqref{eq:CLT-SD}. According to the central limit theorem for independent and identically distributed random vectors, the variance matrix of $\mathbf{Z}$ corresponds to the variance of any of the summands in Eq.~\eqref{eq:CLT-SD}. If we denote by $\mathbf{\Sigma}_{n,m}$ the $(n,m)$th entry of $\mathbf{\Sigma}$ (for $ 1\leq n \leq d$ and $ 1\leq m \leq d$) , we have that the components of the variance matrix $\mathbf{\Sigma}$ can be written as,
\begin{equation}
\label{eq:int-var-rs}
(\mathbf{\Sigma})_{n,m} := \langle (\Delta {\mathbf{r}} +\Delta \tilde {\mathbf{r}} )_{n}  (\Delta {\mathbf{r}} +\Delta \tilde {\mathbf{r}} )_{m}\rangle,
\end{equation}
where we used the fact that $ \Delta {\mathbf{r}}$ and $ \Delta \tilde {\mathbf{r}} $ have both zero mean. In Eq.~\eqref{eq:int-var-rs} $ (\Delta {\mathbf{r}} +\Delta \tilde {\mathbf{r}} )_{n}$ and $ (\Delta {\mathbf{r}} +\Delta \tilde {\mathbf{r}} )_{m}$ stands for the $n$th and $m$th components of $ \Delta {\mathbf{r}} +\Delta \tilde {\mathbf{r}}$, respectively. Performing some calculations we obtain
\begin{eqnarray}
(\mathbf{\Sigma})_{n,m} &=& \langle (\Delta {\mathbf{r}} )_{n}  (\Delta {\mathbf{r}} )_{m}\rangle + 
\langle (\Delta \tilde {\mathbf{r}} )_{n}  (\Delta \tilde {\mathbf{r}} )_{m}\rangle,
\end{eqnarray}
where we used the fact that $\Delta {\mathbf{r}}$ and $\Delta \tilde{\mathbf{r}}$ are independent and have zero mean, which implies that $ \langle \Delta {\mathbf{r}} \cdot \Delta \tilde{\mathbf{r}} \rangle = \mathbf{0}$. Now, using the fact that $\Delta {\mathbf{r}} = \ell_j \mathbf{e}_j$ we see that
\begin{eqnarray}
(\mathbf{\Sigma})_{n,m} &=& \langle \ell_j^2 \rangle \langle (\mathbf{e}_j )_{n}  (\mathbf{e}_j)_{m}\rangle + 
\langle (\Delta \tilde {\mathbf{r}} )_{n}  (\Delta \tilde {\mathbf{r}} )_{m}\rangle,
\end{eqnarray}
Which shows that the variance matrix can be put in terms of the underlying distributions of the model. 

Finally we compute the expected value of $|\mathbf{Z}|^2$. To this end, we use the fact that the moments of $\mathbf{Z}$ are related to the moments of $ \Delta {\mathbf{r}}  + \Delta \tilde {\mathbf{r}} $. Specifically, we can write
\begin{equation}
\langle |\mathbf{Z}|^2 \rangle =  \langle \big| \Delta  {\mathbf{r}} + \Delta \tilde {\mathbf{r}} \big|^2 \rangle = \langle \big| \Delta  {\mathbf{r}}  \big|^2 \rangle +   \langle \big| \Delta \tilde {\mathbf{r}} \big|^2 \rangle,
\end{equation}
where we used the fact that $\langle \Delta  {\mathbf{r}} \cdot \Delta \tilde {\mathbf{r}}  \rangle  = 0$. Next, we should notice that
\begin{eqnarray}
\langle \big| \Delta \tilde {\mathbf{r}} \big|^2 \rangle &=&  2a^2,
\end{eqnarray}
where we used the result given in Eq.~\eqref{eq:sg_tilder}. On the other hand, since $\ell_j$ is distributed according to the free path distribution (which an exponential distribution) it is clear that
\begin{eqnarray}
\langle \big| \Delta  {\mathbf{r}}  \big|^2 \rangle &=&   \langle \ell_j^2 \rangle = 2\lambda^2.
\end{eqnarray}
These results allows us to see that 
\begin{equation}
\langle |\mathbf{Z}|^2 \rangle =  2\lambda^2+  2a^2,
\end{equation}
which is the result we anticipated in Eq.~\ref{eq:Z2}.

\section{Square displacement for power-law distribution of random adsorption time}
\label{ape:2}

In this Appendix we will compute the asymptotic behavior of the mean square displacement of an active particle when the PDF of the adsorption time is a power law as stated in Eq.~\eqref{eq:tau-powlaw},
\begin{equation}
\label{eq:tau-powlaw-ape}
\rho_\tau(t) = \left\{ \begin{array} 
            {r@{\quad \mbox{ if } \quad}l} 
\alpha t^{-\alpha -1}    &  t \geq 1  \\ 
0    & t<1.
             \end{array} \right.
\end{equation}

We will consider the cases $i$) $0 < \alpha < 1$, $ii$) $1 < \alpha < 1$, $iii$) $\alpha =1$ and $iv $) $\alpha = 2$ separately.

Before start to exploring these cases separately, we recall that, in order to obtain the asymptotic properties of the trajectory of an active particle we have to obtain an expression for the net displacement $ \mathbf{R}(t) $, explicitly  as a function of $t$. The starting point is the equation for the net displacement in terms of the number of collisions (actually as a sum of random vectors),
\begin{equation}
\label{eq:displacement-sum-ape}
 \mathbf{R}(t) = \sum_{j=0}^{n} \left( \Delta \mathbf{r}_j  + \Delta \tilde{\mathbf{r}}_j \right),
\end{equation}
and the total time $t$ as well in terms of the number of collisions
\begin{equation}
\label{eq:time-disp-sum-ape}
t = \sum_{j=1}^n \Delta t_j + \sum_{j=1}^n \tau_j,
\end{equation}
It is clear that changing the statistical properties of the adsorption times does not affect the statistical properties of the free path between collisions. Therefore it is clear that we can still made use of the approximation provided by the central limit theorem for the net displacement given in Eq.~\eqref{eq:sumD-approx-ape}, 
\begin{equation}
\label{eq:sumD-approx-ape}
\mathbf{R}(t)  = \sum_{j=0}^{n}  \left( \Delta \mathbf{r}_j  + \Delta \tilde{\mathbf{r}}_j \right)\approx \sqrt{n} \mathbf{Z}, \qquad \mbox{for} \quad n\gg 1,
\end{equation}
where $\mathbf{Z}$ is a multivariate random normal vector with zero mean and variance matrix $\mathbf{\Sigma}$.

\subsection{Case:  $0 < \alpha < 1$}

In this case the sum of random adsorption times, 
\begin{equation}
 \sum_{j=1}^n \tau_j,
\end{equation}
converges to a stable law if normalized according to,
\begin{equation}
\lim_{n\to \infty} \frac{ \sum_{j=1}^n \tau_j }{n^{1/\alpha}} = W,
\end{equation}
where $W$ is a $\alpha$-stable distribution~\cite{gnedenko1954limit}.  According to this result, it is clear that
\begin{equation}
\label{eq:taus}
  \sum_{j=1}^n \tau_j  \approx W n^{1/\alpha},
\end{equation}
for large $n$. On the other hand, the sum 
\begin{equation}
\sum_{j=1}^n \Delta t_j 
\end{equation}
still converges to a normal distribution because it is related to the sum of free path between collisions as follows
\begin{equation}
\sum_{j=1}^n \Delta t_j  = \frac{1}{v}\sum_{j=1}^n \ell_j.
\end{equation}
This implies that
\begin{equation}
\lim_{n\to \infty} \frac{ \sum_{j=1}^n \Delta t_j  - n \lambda /v}{ n^{1/2} (\lambda/v)^2 } = Z_0
\end{equation}
where $Z_0$ has a standard normal distribution. Therefore we can made the following approximation
\begin{equation}
\label{eq:Dts}
\sum_{j=1}^n \Delta t_j \approx  \left( \frac{\lambda}{v} \right) n +  \left( \frac{\lambda}{v} \right) n^{1/2} Z_0,
\end{equation}
for large $n$.
Then, using approximations~\eqref{eq:taus} and~\eqref{eq:Dts}, we have that
\begin{eqnarray}
\sum_{j=1}^n \Delta t_j + \sum_{j=1}^n \tau_j &\approx&   \left( \frac{\lambda}{v} \right) n +  \left( \frac{\lambda}{v} \right) n^{1/2} Z_0 + W n^{1/\alpha}
\nonumber
\\
&\approx& 
 W n^{1/\alpha}.
\end{eqnarray}
In the last equality we used the fact that $0< \alpha < 1$, which implies that $n^{1/\alpha}$ is the leading order in the above equation. Thus, the number of collisions $n$ occurring up to time $t$ can be obtained (for large $t$) from the following equation
\begin{equation}
\label{eq:t-n-0a1}
t \approx  W n^{1/\alpha}.
\end{equation}
Solving for $n$ we obtain,
\begin{equation}
 n \approx \frac{t^{\alpha}}{W^{\alpha}}.
\end{equation}
Using this result into Eq.~\eqref{eq:sumD-approx-ape}, we can see that
\begin{equation}
\label{eq:sumD-int2}
\mathbf{R}(t)   \approx t^{\alpha/2}  \frac{\mathbf{Z}} {W^{\alpha/2}},
\end{equation}
for asymptotically large $t$. Therefore we have that the square displacement goes as
\begin{equation}
\label{eq:sqr-fluct}
\left| \mathbf{R}(t)  \right|^2 \approx t^{\alpha}  \frac{|\mathbf{Z}|^2} {W^{\alpha}}. 
\end{equation}
or equivalently, the mean square displacement behaves as
\begin{equation}
\label{eq:mean-sqr-fluct}
\langle \left| \mathbf{R}(t)   \right|^2 \rangle - \big| \langle  \mathbf{R}(t)   \rangle \big|^2  \propto t^{\alpha},
\end{equation}
for asymptotically large $t$.
This result states that the square displacement grows in time lower than linear with an exponent $\beta = \alpha$, the same as the one for the distribution of the random adsorption times. Notice that this behavior does not depend on the details of the disorder.

\subsection{Case:  $1 < \alpha < 2$}

For $1 < \alpha < 2$ the mean adsorption time $\bar{\tau}$ is finite, but the sum of random adsorption times 
\begin{equation}
\label{eq:sum-ape-1a2}
\sum_{j=1}^n \tau_j,
\end{equation}
does not satisfy the central limit theorem because the variance of $\tau_j$ is not finite. However, if normalized appropriately, this sum converges to a stable law. Specifically we have that
\begin{equation}
\lim_{n\to \infty} \frac{ \sum_{j=1}^n \tau_j - n\bar{\tau} }{n^{1/\alpha}} = W,
\end{equation}
where $W$ is a $\alpha$-stable distribution~\cite{gnedenko1954limit}. This expression allows us to approximate the sum~\eqref{eq:sum-ape-1a2} as follows,
\begin{equation}
\label{eq:tau-1a2}
\sum_{j=1}^n \tau_j  \approx n\bar{\tau} + W n^{1/\alpha}, 
\end{equation}
for $n\gg 1$.
Using the approximation given in~\eqref{eq:Dts} for the sum of $\Delta t_j$, and the above expression for the sum of $\tau_j$, we have that
\begin{eqnarray}
t &=& \sum_{j=1}^n \tau_j + \sum_{j=1}^n \Delta t_j 
\nonumber
\\
&\approx&
n\bar{\tau} + W n^{1/\alpha} + \left( \frac{\lambda}{v} \right) n + \left( \frac{\lambda}{v} \right) n^{1/2} Z_0
\nonumber
\\
&=&\left( \bar{\tau} +\frac{\lambda}{v} \right) n +  W n^{1/\alpha} +  \left( \frac{\lambda}{v} \right) n^{1/2} Z_0.
\label{eq:ts-1a2}
\end{eqnarray}
Since $1 < \alpha <2$, it is clear that $ 1/2 < \alpha^{-1} <1$, which means that the leading order in the above expression is $n$. Therefor we have that
\begin{equation}
t \approx \left( \bar{\tau} +\frac{\lambda}{v} \right) n,
\end{equation}
for $n\gg 1$. Equivalently we have that 
\begin{equation}
n \approx \frac{t}{ \bar{\tau} +\frac{\lambda}{v} }.
\end{equation}
Using this last result in Eq.~\eqref{eq:sumD-approx-ape} we obtain for the net displacement,
\begin{equation}
\mathbf{R}(t)    \approx  \left(  \frac{t}{ \bar{\tau} +\frac{\lambda}{v} } \right)^{1/2}  \mathbf{Z}.
\end{equation}
Then, the mean square displacement can be computed from the last relation, obtaining,
\begin{equation}
\langle \left| \mathbf{R}(t)   \right|^2 \rangle - \big| \langle  \mathbf{R}(t)   \rangle \big|^2   \approx  \left(  \frac{t}{ \bar{\tau} +\frac{\lambda}{v} } \right)   \langle |\mathbf{Z}|^2 \rangle.
\end{equation}
 Since $\langle |\mathbf{Z}|^2 \rangle = 2\lambda^2 + 2a^2$ it is clear that
\begin{equation}
D_{\mathrm{eff}} = 
\frac{1}{d}\left(  \frac{\lambda^2 + a^2}{ \bar{\tau} +\frac{\lambda}{v} } \right) ,
\end{equation}
which is the expression we have already obtained for the diffusion coefficient when the central limit theorem takes place.

%
%

\subsection{Case:  $\alpha =1$ }

For $\alpha  =1$ the mean adsorption time $\bar{\tau}$ is infinite (as well as its variance) and the sum of random adsorption times converges to a stable law in the following form 
\begin{equation}
\lim_{n\to \infty} \frac{ \sum_{j=1}^n \tau_j  }{n\ln(n)} = W,
\end{equation}
where $W$ is a $\alpha$-stable distribution for $\alpha = 1$~\cite{gnedenko1954limit}.
This limit theorem allows us to approximate the sum of random adsorption times  as follows
\begin{equation}
\label{eq:taus-a1}
\sum_{j=1}^n \tau_j  =  W n\ln(n).
\end{equation}
This expression together with Eq.~\eqref{eq:Dts} allows us to write
\begin{eqnarray}
t &=& \sum_{j=1}^n \tau_j + \sum_{j=1}^n \Delta t_j 
\nonumber
\\
&\approx&   W n\ln(n) + \left( \frac{\lambda}{v} \right) n + \left( \frac{\lambda}{v} \right) n^{1/2} Z_0. 
\label{eq:t-a1}
\end{eqnarray}
As we can see the leading order in the above approximation is $n\ln(n)$. In Ref.~\cite{Hidalgo2017Scarce} it has been proved that $n$ as function of $t$ can be written  approximately as,
\begin{equation}
n \approx \frac{t/W}{\ln(t/W)},
\end{equation}
for large $t$. 
Using this result in Eq.~\eqref{eq:sumD-approx-ape} we obtain
\begin{equation}
\mathbf{R}(t)  \approx  \left(  \frac{t/W}{\ln(t/W)} \right)^{1/2}  \mathbf{Z}.
\end{equation}
Then, the square displacement can be written as, 
\begin{equation}
 \left| \mathbf{R}(t) \right|^2   \approx  \left( \frac{t/W}{\ln(t/W)} \right)   |\mathbf{Z}|^2.
\end{equation}
This result implies that the mean square displacement grows in time as
\begin{equation}
\langle \left| \mathbf{R}(t)   \right|^2 \rangle - \big| \langle  \mathbf{R}(t)   \rangle \big|^2     \propto   \frac{t}{\ln(t)},
\end{equation}
which is a marginal subdiffusion because the growth is lower than linear but not a power law.

\subsection{Case: $\alpha = 2$}
For $\alpha  =2$ the mean adsorption time $\bar{\tau}$ is finite but its variance is infinite.   The sum of random adsorption times converges to a normal distribution in the following form 
\begin{equation}
\lim_{n\to \infty} \frac{ \sum_{j=1}^n \tau_j  - n\bar{\tau}}{ \sqrt{n\ln(n)}} = W,
\end{equation}
where $W$ is a normally distributed random variable with zero mean~\cite{gnedenko1954limit}. This result allows us to write the sum  of random adsorption times approximately as.
\begin{equation}
\sum_{j=1}^n \tau_j  \approx  n\bar{\tau} + \sqrt{n\ln(n)}  W,
\end{equation}
This expression together with Eq.~\eqref{eq:Dts} allows us to write
\begin{eqnarray}
t &=& \sum_{j=1}^n \tau_j + \sum_{j=1}^n \Delta t_j 
\nonumber
\\
&\approx&    n\bar{\tau} + \sqrt{n\ln(n)}  W + \left( \frac{\lambda}{v} \right) n + \left( \frac{\lambda}{v} \right) n^{1/2} Z_0
\nonumber
\\
&=&    \left( \bar{\tau}  + \frac{\lambda}{v} \right) n  + \sqrt{n\ln(n)}  W + \left( \frac{\lambda}{v} \right) n^{1/2} Z_0. 
\end{eqnarray}
As we can see the two leading orders in the above approximation are $n$ and $\sqrt{n\ln(n)} $. Retaining the two first leading orders in the above approximation, we have that $n$ as function of $t$ can be written  approximately as (see Appendix B.3 in Ref.~\cite{Hidalgo2017Scarce}),
\begin{equation}
n \approx \frac{t}{  \bar{\tau}  + \frac{\lambda}{v}} + O\left( (t\ln (t))^{1/2}\right ).
\end{equation}
Thus, the square displacement can be written as, 
\begin{equation}
 \left| \mathbf{R}(t) \right|^2   \approx  \left( \frac{t}{  \bar{\tau}  + \frac{\lambda}{v}}  \right)   |\mathbf{Z}|^2.
\end{equation}
This result implies that the square displacement grows linearly in time, i.e.,
\begin{equation}
\langle \left| \mathbf{R}(t)   \right|^2 \rangle - \big| \langle  \mathbf{R}(t)   \rangle \big|^2   \approx  \left( \frac{t}{  \bar{\tau}  + \frac{\lambda}{v}}  \right)  \langle |\mathbf{Z} |^2 \rangle.
\end{equation}
The latter means that the effective diffusion coefficient is finite and results in 
\begin{equation}
D_{\mathrm{eff}} = 
\frac{1}{d}\left(  \frac{\lambda^2 + a^2}{ \bar{\tau} +\frac{\lambda}{v} } \right) ,
\end{equation}
which is the expression we have already obtained for the diffusion coefficient when the central limit theorem takes place.

\section{Fraction of adsorbed particles}
\label{ape:3}

In this Appendix we compute the fraction of adsorbed particles using the heavy-tailed distribution model for the random adsorption times. According to Eq.~\eqref{eq:gamma_taus} the fraction of adsorbed particles $\gamma $ can be written as
\begin{equation}
\gamma = \lim_{t\to \infty} \frac{1}{t} \sum_{j=1}^n \tau_j.
\end{equation}

As we saw in Appendix~\ref{ape:2} we have that, for $ 0 <\alpha < 1$ the sum of random adsorption times can be approximated as (see Eq.~\eqref{eq:taus}), 
\begin{equation}
\sum_{j=1}^n \tau_j  \approx W n^{1/\alpha},
\end{equation}
On the other hand, according to Eq.~\eqref{eq:t-n-0a1}, we can write $t$ as 
\begin{equation}
t \approx  W n^{1/\alpha},
\end{equation}
for $ 0 <\alpha < 1$ and large $t$. Thus, in the limit $t\to  \infty$ it is clear that
\begin{equation}
\gamma = \lim_{t\to \infty} \frac{1}{t} \sum_{j=1}^n \tau_j 
= \lim_{n \to \infty} \frac{1}{ W n^{1/\alpha} } W n^{1/\alpha} = 1,
\end{equation}
which proves that, in the average, all the particles get adsorbed  for $ 0 <\alpha < 1$. 

For $\alpha = 1$, a similar behavior occurs. Looking at Eqs.~\eqref{eq:taus-a1} and~\eqref{eq:t-a1} , we can see that,
\begin{eqnarray}
\sum_{j=1}^n \tau_j   &\approx&  W n\ln(n),
\\  
t  &\approx&   W n\ln(n),
\end{eqnarray}
for $t$ going to infinity. Thus it is clear that 
\begin{equation}
\gamma  = \lim_{n \to \infty} \frac{1}{ W n\ln(n) } W n\ln(n) = 1,
\end{equation}
for $\alpha = 1$. 

Next, for the case $1 < \alpha < 2$, Eqs.~\eqref{eq:tau-1a2}and~\eqref{eq:ts-1a2} gives us approximations for $t$ and $\sum_{j=1}^n \tau_j$ for large $t$, 
\begin{eqnarray}
\label{eq:taus-gamma}
\sum_{j=1}^n \tau_j  &\approx&  n\bar{\tau} 
\\
t &\approx&\left( \bar{\tau} +\frac{\lambda}{v} \right) n.
\label{eq:ts-gamma}
\end{eqnarray}
These two approximations allows us to compute $\gamma$ as follows,
\begin{equation}
\gamma  = \lim_{n \to \infty} \frac{1}{ \left( \bar{\tau} +\frac{\lambda}{v} \right) n  }   n\bar{\tau} + W n^{1/\alpha} 
= \frac{\bar{\tau} }{  \bar{\tau} +\frac{\lambda}{v}   } ,
\end{equation}
for the case $1 < \alpha < 2$. It is not hard to see that for the case $\alpha \geq 2$ both, $t$ and $ \sum_{j=1}^n \tau_j$ grow linearly in time (because for the latter, the central limit theorem is valid), and actually  they can be approximated in the same way as shown in Eqs.~\eqref{eq:taus-gamma} and~\eqref{eq:ts-gamma}. Then we can extend the last result to state that the fraction of adsorbed  particles is given by 
\begin{equation}
\gamma  =  \frac{\bar{\tau} }{  \bar{\tau} +\frac{\lambda}{v}   } ,
\end{equation}
for the case $\alpha \geq 2$ and, in general, when the central limit theorem takes place.

%
%

%
%

\nocite{*}

\bibliography{Active_refs}

\end{document}